\documentclass[12pt]{article}

\usepackage{moreverb}
\usepackage{booktabs}

\usepackage[colorlinks,bookmarksopen,bookmarksnumbered,citecolor=red,urlcolor=red]{hyperref} 
\usepackage{tikz,xcolor,amsmath, amssymb, amsfonts, amsthm}
\definecolor{orcidlogocol}{HTML}{A6CE39}
\RequirePackage{bm}
\usepackage[ruled,vlined]{algorithm2e}

\usepackage{natbib}

\newcommand\BibTeX{{\rmfamily B\kern-.05em \textsc{i\kern-.025em b}\kern-.08em
T\kern-.1667em\lower.7ex\hbox{E}\kern-.125emX}}

\def\volumeyear{2012}

\newcommand{\xb}{\mathbf{x}}
\newcommand{\yb}{\mathbf{y}}
\newcommand{\zb}{\mathbf{z}}

\newcommand{\Zb}{\mathbf{Z}}

\newcommand{\bX}{\bm{X}}

\newcommand{\bZ}{\bm{Z}}

\DeclareMathOperator*{\argmin}{argmin}

\newcommand{\RN}[1]{%
	\textup{\uppercase\expandafter{\romannumeral#1}}%
}

\theoremstyle{plain}
\newtheorem{theorem}{Theorem}

\theoremstyle{definition}

\theoremstyle{remark}

\begin{document}

\title{Deep fiducial inference}

\author{Gang Li and Jan Hannig\thanks{
	jan.hannig@unc.edu}\hspace{.2cm} \\
\\
Department of Statstics and Operations Research,\\ University of North Carolina at Chapel Hill, 
NC 27599-3260, USA\\}

\maketitle




\begin{abstract}
Since the mid-2000s, there has been a resurrection of interest in modern modifications of fiducial inference. To date, the main computational tool to extract a generalized fiducial distribution is Markov chain Monte Carlo (MCMC). We propose an alternative way of computing a generalized fiducial distribution that could be used in complex situations.  In particular, to overcome the difficulty when the unnormalized fiducial density (needed for MCMC), we design a fiducial autoencoder (FAE). The fitted autoencoder is used to generate generalized fiducial samples of the unknown parameters. To increase accuracy, we then apply an approximate fiducial computation (AFC) algorithm, by rejecting samples that when plugged into a decoder do not replicate the observed data well enough. Our numerical experiments show the effectiveness of our FAE-based inverse solution and the excellent coverage performance of the AFC corrected FAE solution.
\end{abstract}

\noindent%
{\it \textbf{Keywords}:}  Deep Fiducial Inference, Fiducial Autoencoder, Approximate Fiducial Inference, Uncertainty Quantification


\section{Introduction}\label{sec1}

Generalized fiducial inference (GFI) \cite{hannig2016generalized}, a modern re-incarnation of R. A. Fisher's fiducial inference \citep{fisher_1930}, provides inferentially meaningful probability statements about subsets of parameter space without the need for subjective prior information. GFI specifies a generalized fiducial distribution (GFD)  by defining a data-dependent measure on the parameter space through an inverse of a data-generating algorithm (see Sections~\ref{sec2}). Data-generating algorithm plays the role of a model and is sometimes called data-generating equation or data-generating function. With GFD as a distribution estimator for the fixed parameter, we can further define approximate confidence (fiducial) sets which are often shown in simulation to have very desired properties. 

Given the {data-generating algorithm} and the corresponding density of the GFD, one could form point estimate and asymptotic confidence sets similarly as with a Bayesian posterior density. Standard MCMC-type sampling techniques have already been successfully implemented in many situations; see \cite{hannig2016generalized} and the references therein. However, sometimes the generalized fiducial density can be hard to compute. Especially, when the likelihood of the data is intractable, MCMC would be difficult or impossible to implement. 
For this reason, we propose using a deep neural network to approximate the nonlinear inverse to the data-generating algorithm and to generate an approximation to the fiducial distribution without knowing the exact form the density.

Autoencoder (AE) \cite{hinton1994autoencoders,schmidhuber2015deep} is a type of neural network architecture to learn data code and to reduce data dimensions. Different variants of autoencoder \cite{vincent2010stacked} have gained a lot of successful results in applications of the neural network, such as variational autoencoder \cite{doersch2016tutorial} and variational Bayes \cite{kingma2013auto}.  Besides, the denoising autoencoder \cite{vincent2008extracting} is designed to remove the noise effects without losing the main information. Inspired by the autoencoder's architecture, we design a fiducial autoencoder (FAE), which continues to use the neural network as the encoder to approximate the inverse of data generating function but employ the exact data-generating function as the decoder. Furthermore, we designed and implemented approximate fiducial computation algorithms to generate generalized fiducial samples. 


The rest of this paper is organized as follows. In Section~\ref{sec2}, we first introduce the generalized fiducial inference and its standard MCMC based solution. In Section~\ref{sec3}, we design a fiducial autoencoder (FAE) to approximate the inverse function when it is not directly available, and we propose the approximate fiducial computation (AFC) algorithm {to further increase the accuracy of FAE}. We demonstrate the performance
of FAE and AFC in three numerical examples in Section~\ref{sec4}. Section~\ref{sec5} concludes the paper with some discussions. 


\section{Background on Generalized Fiducial Inference}\label{sec2}
Before introducing our computational tool, fiducial autoencoder (FAE), we first briefly present the current state of ideas in generalized fiducial inference (GFI). The GFI framework is based on linking the observed data $\xb$, the unknown parameter $\mu$, and some random component $\zb$ via a data generating algorithm, also called data generating algorithm. We shall discuss this in detail.

\noindent
\textbf{Data generating algorithm:} The data generating algorithm is
\begin{equation}\label{eq:dgf}
	\xb=f(\zb,{\mu}).
\end{equation}
where $\xb$ is the data,  ${\mu}$ is the parameter, and $\zb$ is a random component with distribution $F_0$ that is completely known and independent of the parameter $\mu$. It is assumed that the data could have been generated by fixing some parameter value $\mu$, generating a value $\zb$ from distribution $F_0$, and plugging them into the equation \eqref{eq:dgf}. The data $\xb$ is assumed observed, while the values of $\mu$ and $\zb$ are unobserved. Notice that this procedure uniquely determines the sampling distribution of $\xb$.

\noindent
\textbf{Generalized fiducial distribution:}
If both $\xb$ and $\zb$ were known, then inverting equation \eqref{eq:dgf}, i.e., solving for $\mu$, would give us the unknown parameter. Since $\zb$ is unknown, we estimate it using its distribution $F_0$. Generalized fiducial distribution (GFD) is then defined by propagating the distribution $F_0$ through the inverse of the data generating algorithm. 
This idea is refined in the following rigorous mathematical definition \citep{hannig2016generalized}:

For an $\epsilon>0$, consider the following inverse problem
\begin{equation}\label{eq:FIDopt}
 g(\xb,\zb)=\argmin_{\mu^{\star}} {\| \xb - f(\zb, \mu^{\star} ) \|}.
\end{equation}
In this paper we will use $\|\cdot\|$ as either $\ell_2$ or Frobenius norm and call the  inverse of the data generating algorithm $g(\xb,\zb)$ the inverse function.  
Next define the random variable $\mu_\epsilon^\star=g(\xb,\Zb_\epsilon^{\star})$,
where $\Zb_\epsilon^{\star}$ has distribution $F_0$ truncated to the set  
\begin{equation}\label{eq:truncate}
\mathcal C_\epsilon = \{ \Zb_\epsilon^{\star} : \| \xb - f(\Zb_\epsilon^{\star}, \mu_\epsilon^\star ) \| = \| \xb - f(\Zb_\epsilon^{\star}, g(\xb,\Zb_\epsilon^{\star}) ) \| \leq \epsilon \}.
\end{equation}
Then assuming that the random variables $\mu_\epsilon^\star$ converge in distribution as $\epsilon\to 0$, GFD is defined as the limiting distribution $\mu^\star= \lim_{\epsilon\to 0} \mu_\epsilon^\star$. Notice that the fiducial distribution of $\mu^\star$ depends on the observed data $\xb$. 

With this GFD as the distribution estimator for $\mu$, one can form point estimators and construct approximate confidence sets just like using a Bayesian posterior distribution. Notice that GFI does not need any prior information, and in fact, it provides a systematic approach to deriving objective Bayes-like posterior distributions \citep{hannig2016generalized}. 

The following theorem is a basis for most current numerical implementations of GFD: 
\begin{theorem}[\cite{hannig2016generalized}]
	\label{Jacobian}
	Under mild condition, the limiting distribution above has a density 
	\[r_{\xb} ( \boldsymbol { \mu } ) = \frac { L ( \xb | \boldsymbol { \mu } ) J ( \xb , \boldsymbol { \mu } ) } { \int  L \left( \xb  | \boldsymbol { \mu } ^ { \prime } \right) J \left( \xb , \boldsymbol { \mu } ^ { \prime } \right) d \boldsymbol { \mu } ^ { \prime } }\]
	where $L ( \xb | \boldsymbol { \mu } )$ is the likelihood function and
	$J ( \xb , \boldsymbol { \mu } ) = D \left( { \nabla_{ \boldsymbol { \mu } } } f \left. ( \zb , \boldsymbol { \mu } ) \right| _ { \zb = f ^ { - 1 } ( \xb , \boldsymbol { \mu } ) } \right),$
	where $\nabla_{ \boldsymbol { \mu }} $ is a gradient matrix with respect to $\mu$ and $D ( A ) = \left( \operatorname { det } A ^ { \prime } A \right)^{ \frac { 1 } { 2 } }$. 
\end{theorem}

Notice that the form of the GFD density in the theorem has a similar form to a Bayesian's posterior density, where Jacobian function $J(x,\mu)$ plays the role of a data-dependent prior.
Similarly to Bayesian inference, one potential challenge using this formula to calculate the fiducial density is that the denominator might be intractable. Traditionally this has been addressed using MCMC algorithms.  There are many well-known challenges associated with MCMC such as computational speeds and intractable high dimensional integrals. In particular, MCMC is impossible to implement when a closed form of the likelihood is unknown or difficult to derive. 

 In this paper we propose a sampling-based computational solution using \eqref{eq:FIDopt} directly. There are two main challenges in implementing this approach. First, the solution to the optimization problem in \eqref{eq:FIDopt} might not have an analytical form or might be difficult to calculate. Therefore we propose to approximate the solution  $g(\xb,\Zb_\epsilon^{\star})$ using a deep neural network (see details in section \ref{dfi:dfi}). Second, one needs to generate $\Zb_\epsilon^{\star}$ from $\mathcal C_\epsilon$ for a small $\epsilon$, which could be very challenging when $\mathcal C_0$ is a lower dimensional-manifold. We propose an approximate fiducial computation (AFC) algorithm (see details in section \ref{dfi:afc}) to effectively generate fiducial samples. 

\noindent
\textbf{Confidence Curve:}
Before proceeding further, we also introduce the confidence curve (CC), a useful graphical tool for plotting epistemic distributions \citep{birnbaum1961foundations}. Given $ R_\xb( \mu )$ the distribution function of GFD corresponding to a one-dimensional marginal of density in Theorem~\ref{Jacobian}, CC is defined as $2 | R _\xb( \mu ) - 0.5 |$. CC shows two-sided confidence intervals at all significance levels stacked upon each other. The median of GFD is the point where CC touches $x$-axis and can be used as a point estimator. 
%

\section{Methodology}\label{sec3}

\subsection{Fiducial Autoencoder}\label{dfi:dfi}
In this section, we design a fiducial autoencoder (FAE), a deep neural network based approximation for the inverse function \eqref{eq:FIDopt}, for circumstances where the inverse function might not have an analytical form or might be difficult to calculate. There are standard theoretical guarantees for a good approximation \citep{hornik1989multilayer}  and our numerical experiments in Section~\ref{sec3} also validate this approach.
\begin{theorem}[Universal Approximation Theorem] \label{thm2} A feedforward network with a linear output layer and at least one hidden layer with any “squashing” activation functions can approximate any Borel measurable function, provided that the network is given enough hidden units.
\end{theorem}
\begin{figure}[b]
	\centering
	\includegraphics[width=0.8\linewidth ,height=1.4in]{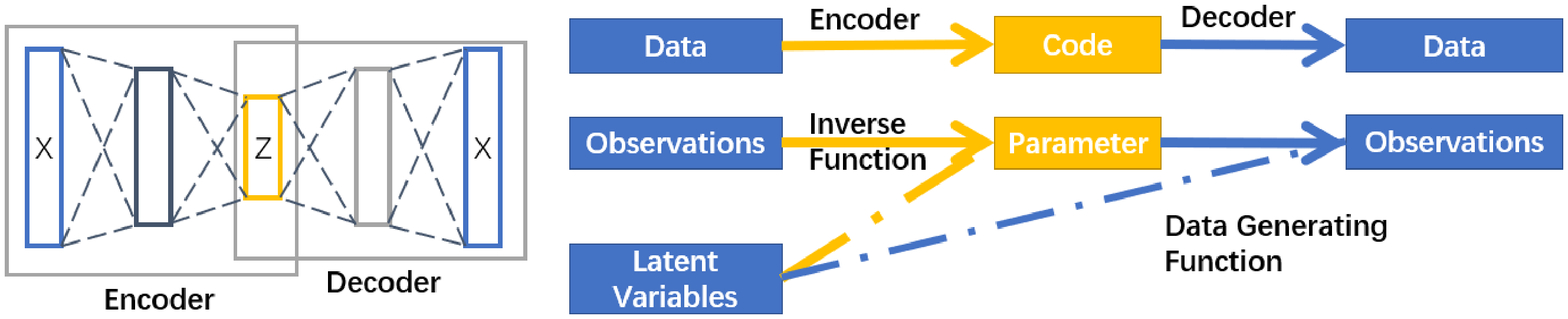}
	\caption{\label{Fig:AE} The left panel shows a schematic of a generic autoencoder while the right panel compares autoencoder with usual generalized fiducial inference.}
\end{figure}

The most basic version of autoencoder (AE) \citep{hinton1994autoencoders,schmidhuber2015deep} often contains two parts, encoder and decoder. The encoder maps from the observation space $\mathcal{X}$ to latent coder space $\mathcal{Z}$ while the decoder does the inverse (see left panel of Figure \ref{Fig:AE}).
We compare the encode-decode process of the standard AE to the usual fiducial inference in the right panel of Figure \ref{Fig:AE}. The process by which the data generating function maps the parameter and the latent variables to the observations is similar to the mechanism by which the decoder maps the code to the data in the usual AE. Similarly, the encoder in AE plays a role akin to the inverse of the data generating algorithm in fiducial inference.
Inspired by this analogy, we create the FAE framework which will use as an encoder a neural network that will approximate the inverse of the data generating algorithm $\hat\mu=g(\xb,\zb)$ to be estimated, and as  a decoder the data generating algorithm $\hat{\xb}=f(\zb,\hat\mu)$ that is known to us.

\begin{figure}[h]
	\centering
	\includegraphics[width=0.5\linewidth]{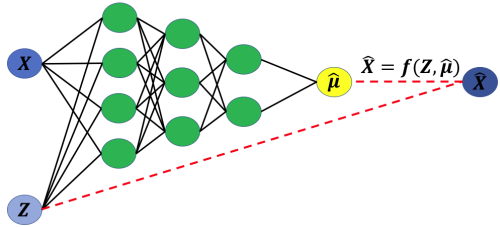}
	\caption{\label{Fig:FAE} Overview of standard fiducial autoencoder architecture. The encoder is a neural network to be fitted and decoder is the known data-generating algorithm; $\bZ$ is used in both encoder and decoder.}
\end{figure}

More precisely, the standard FAE has two input nodes, the observation data $\xb$ and the latent variable $\zb$, one final output node, the prediction $\hat{\xb}$, and one intermediate output node, the prediction $\hat{\mu}$. The FAE's encoder {usually} consists of a deep fully connected neural network, which maps the $\xb$ and $\zb$ to $\hat{\mu}$. The architecture of the encoder is flexible and should be selected based on the problem so that it can learn the inverse function well. Unlike the traditional autoencoder which uses the code layer $\hat{\mu}$ to predict the input nodes, the FAE's decoder employs the known data generating algorithm $\hat{\xb}=f(\zb,\hat{\mu})$. We built an additional straight connection from the input $\zb$ node directly to the final output node $\hat{\xb}$, which is an extreme case of the residual neural network, see Figure~\ref{Fig:FAE}. 
{Two main differences between traditional autoencoder and our FAE are summarized as follows:}
\begin{itemize}
	\item First, we use the exact data generating algorithm instead of another group of neural network as the decoder.
	\item Second, our decoder takes as the input not only code layer $\hat{\mu}$ but also the random component $\bZ^\star$.
\end{itemize}

One big advantage of FAE, compared to the usual AE, is that  it never suffers a lack of training data. This is because the training data is obtained by simulation from the data generating algorithm that is known to us. In particular,  given a number of training $\{ \mu_k, k = 1,2,..., n_{train}  \}$, we could generate as many pairs of $\xb_k$ and $\zb_k$ for training FAE as needed. The limitation is that we need to make sure that the training set contains a large enough number of data sets that are similar to the observed data. For example, if the training set does not contain enough values of $\mu_k$ that are close to the true $\mu$, we cannot  guarantee the FAE to provide good answers. Therefore, we recommend that enough of the training data is generated using values of $\mu_k$ close to some pilot estimator of $\hat\mu$.; for example, the least square estimator as in Section~\ref{dfi:s3:BOD}.

Since the decoder is the fixed deterministic data generating function, the loss of FAE quantifies how well the encoder approximates the inverse function. The total loss function we use for training our neural network contains two parts, the mean square error with regard to $\hat{\xb}$ and $\hat{\mu}$:
$L=w_1 {\|\xb-\hat{\xb}\|}^2+w_2 {\| \mu-\hat{\mu} \|}^2,$
where $w_1$ and $w_2$ are user-selected weights.
If we set $w_1=0$ and $w_2=1$, we would in effect be training a neural network mapping directly from $X$ and $Z$ to $\mu$, and we would miss the information provided by the data generating function. Our numerical experiments show that this choice is not optimal. After incorporating the information from the data generating function, the approximation performance is greatly improved. 
On the other hand, if we only use the mean square error based on predicting the data, $\|\xb-\hat{\xb}\|^2$ $(w_1=1$ and $w_2=0)$, the FAE would still do reasonably well. However, using MSE for $\hat{\mu}$ with appropriate weights does not only increase the convergence speed but also improves the FAE's performance. In our numerical experiments we manually select $w_1, w_2$ so that the loss with regard to the observation, $w_1 {\|\xb-\hat{\xb}\|}^2$, and the loss with regard to the parameter, $w_2 {\| \mu-\hat{\mu} \|}^2 $, are roughly of the same magnitude. For example, we set $w_1= w_2 =1$ as the default parameters in Section~\ref{sec:nonlinear}.

\subsection{Approximate Fiducial Computation}\label{dfi:afc}
Once the FAE converges, we simply apply the encoder with pairs of the fixed observed $\xb$ and {a} number of simulated $\bZ^\star$, which are the independent identical copy of $\bZ$, to get the estimated ${\mu^\star}$. Truncating these fiducial samples to the set $\bZ^\star\in\mathcal C_\varepsilon$, defined in \eqref{eq:truncate}, using approximate fiducial computation algorithms, one will achieve samples from the approximate generalized fiducial distribution that can be used for inference. Notice that using only $\bZ^\star\in\mathcal C_\varepsilon$  is very natural as we are discarding those $\bZ^\star$ for which FAE predicts values of $\xb^\star$ that are far from the observed value $\xb$.

This is similar to the approximate Bayesian computation (ABC) that  have been intensively studied in the past years \citep{blum2013comparative}. Similar to ABC, which is designed to overcome the intractable likelihood function, AFC avoids directly calculating the fiducial density and the inverse function. One major difference is that while ABC uses a prior distribution to get candidate $\mu^\star$, the AFC algorithm uses the optimization problem \eqref{eq:FIDopt} eschewing the need to select an arbitrary prior distribution.

The steps of the AFC algorithms are summarized in Algorithm~\ref{alg1}.
Note that for different problems and even for different parameters within the same problem we might need to select different thresholds to efficiently get valid approximate generalized fiducial samples. If the threshold is too big, then we might get biased samples; if the threshold is too small, it would be very difficult for AFC to generate enough samples passing the threshold condition. In practice, we could use one random batch of samples to approximate the distribution of $dist(\xb,{\xb^\star})$, and select the threshold according to the efficiency and the accuracy we expect for the AFC.

We call the samples from Algorithm~\ref{alg1} approximate generalized fiducial samples. Aggregating those samples into an empirical distribution provides an approximation to the generalized Fiducial distribution (GFD) and form the bases for making statistical inference. In our numerical experiments, we study statistical properties  of the point estimators based on the mean and median, and the $\alpha$-level
approximate confidence intervals based on the 
{$(1-\alpha)/2$ and $(1+\alpha)/2$ }
empirical quantiles of the generalized fiducial samples. {The corresponding confidence curves are reported to visualize the approximate fiducial distribution.}
Finally, we remark that the AFC algorithm can be useful even in situations when we know the analytical form of the fiducial inverse function \eqref{eq:FIDopt}. We demonstrate this on 
an example in Section~\ref{sec:Laplace}
\begin{algorithm}[H]
	\SetAlgoLined
	\SetKwInOut{Input}{Input}
	\SetKwInOut{Output}{Output}
	\Input{Data generating function {f}; (approximate) inverse function {g}; distribution $F_0$ for generating $\bZ^\star$; observation $\xb$; {threshold} $\epsilon$}
	\Output{Generalized fiducial samples (GFS)}
	\While{$(itr < max\_itr)$ and  $(\#$ of GFS $< N)$}{
		Sample $\bZ^{\star}$ from $F_0$\;
		${\mu^\star}=g(\xb,\bZ^{\star})$ \;
		${\xb^\star}=f(\bZ^{\star},{\mu^\star})$ \;
		\eIf{$dist(\xb,{\xb^\star}) < \epsilon$ }{
			Accept ${\mu^\star}$\;
		}{
			Reject ${\mu^\star}$\;
		}
		$itr=itr+1$\;
	}
	\caption{\textbf{Approximate Fiducial Computation (AFC)}}\label{alg1}
\end{algorithm}

\section{Numerical experiments}\label{sec4}
We report the results of three numerical experiments. The first is the location-scale Laplace distribution, a relatively straightforward example illustrating AFC when the analytical form of the inverse of the data generating algorithm is available. The second example demonstrates that FAE can learn a non-linear inverse function and shows how AFC improves the inference with the decreasing thresholds $\varepsilon$. 
The third example compares FAE and several competing methods using a highly non-linear regression model. 

\subsection{Location-Scale Laplace Distribution}\label{sec:Laplace}
 Consider the data generating algorithm \[\xb=f(\zb,{\mu})=\theta\times\bm{1}+ \sigma \zb,\] 
 where  $\xb=(x_1,\ldots,x_m)^\top$ are the observed data,  $\mu= (\theta, \sigma)$ are the unknown {scalar} parameters, and $\zb=(z_1,\ldots,z_m)^\top$ is {a vector of i.i.d. samples}  from {standard} Laplace distribution, i.e., density $f(z)={e^{-|z|}}/{2}$.  
Straightforward calculations show that $\mu^\star= (\theta^\star, \sigma^\star) =g(\xb, \bZ^{\star})$ is
\begin{equation}
\sigma^\star =\frac{\sum_{i=1}^m (x_i -\bar x)(Z^{\star}_i-\bar Z^{\star})}{\sum_{i=1}^m (Z^{\star}_i-\bar Z^{\star})^2}
\qquad\mbox{ and }\qquad
\theta^\star = \bar x - \sigma^\star\bar Z^{\star},     
\end{equation}
where $\bar x=m^{-1}\sum_{i=1}^m x_i$ and $\bar Z^{\star}=m^{-1}\sum_{i=1}^m Z^{\star}_i$.

Since we have the analytical solution for the inverse function we can easily compute $\mu^\star=g(\xb,\bZ^\star)$ for any $\bZ^{\star} $ following Laplace distribution. Then we use the data generating algorithm to compute the predicted observation $\xb^\star=f(\bZ^{\star},{\mu}^\star)={\theta}^\star\times\bm{1}+{\sigma}^\star\bZ^{\star}$, and accept the proposed $\mu^\star$ if and only if $\|\xb - {\xb}^\star\|\leq\epsilon$ for some pre-selected threshold $\epsilon$. This process is repeated until we get the desired number of fiducial samples. 

Notice that for small enough $\epsilon$ the AFC algorithm will not accept $\bZ^\star$ that does not match the order of the data. Since the distribution of $\bZ^\star$ is exchangeable, we will without loss of generality assume that both $x_1<\cdots<x_m$ and $Z_1^\star<\cdots<Z_m^\star$ are ordered. This will make the algorithm more efficient. Finally, because of invariance of the location-scale model, it is enough to perform the simulation using only one value the true parameter value, say $\theta=0$ and $\sigma=1$.  

\begin{figure}[t]
	\centering
	\includegraphics[width=0.8\linewidth]{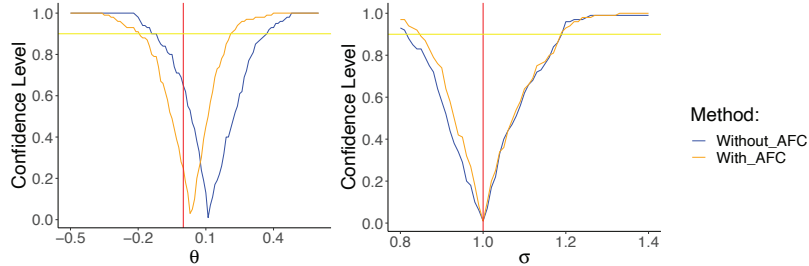}
	\caption{\label{Laplace-CC}  Confidence curves of marginal distributions for the Laplace example.  The red vertical lines shows the true parameters ($\theta=0$ and $\sigma=1$) generated the observed data. The yellow horizontal line indicates the 90\% confidence intervals.}\end{figure}

Figure ~\ref{Laplace-CC} shows an example of CC with and without AFC for the same dataset  containing $m=100$ observations. In the left panel, we see that the CC for $\theta$ with AFC (orange)  is much thinner than the curve without AFC (blue). And the corresponding fiducial median with AFC is also more closed to the truth 0. In the right panel, the CC for $\sigma$ with AFC provides a slightly narrower $90\%$ confidence interval compared with that without AFC.

\begin{center}
	\begin{table}[h]
		\centering
		\caption{ Inference Performance without AFC for Laplace Example.}
		\label{tab:Laplace1}%
		\begin{tabular*}{500pt}{@{\extracolsep\fill}lcccc@{\extracolsep\fill}}
			\toprule
			\textbf{Truth} & \textbf{Coverage} & \textbf{Expected CI Length} & \textbf{Expected Mean} & \textbf{Expected Median} \\
			\midrule
			$\theta=0$ & 0.835 & 0.4468 & 0.0042 & 0.0048 \\ 
			$\sigma=1$ & 0.935 & 0.3636 & 0.9717 & 0.9657 \\
			\bottomrule
		\end{tabular*}
	\end{table}
\end{center}

Tables~\ref{tab:Laplace1} and \ref{tab:Laplace2} compare the inference performance with and without AFC. In particular, we fixed the true parameters $\theta=0$ and $\sigma=1$, and observed 200 data sets each containing $m=100$ observations
$\{ \xb_k=(x_{1,k},\ldots,x_{m,k})^\top,  k = 1, \ldots,200 \}$.  For each $\xb_k$, we used 1,000 random $\{ \bZ_{j,k}^\star, j = 1, \ldots,1000 \}$ to obtain the corresponding $\mu_{j,k}^\star$ and use them to obtain estimators of the mean, median and 90$\%$ confidence set. We report four key statistics averaged over the 200 data sets, namely, coverage, the expected length of the confidence intervals, the expected value of the fiducial mean and median. We can see that AFC provides more accurate point estimations for both mean estimator and median estimator. In addition, the length of confidence intervals with AFC at the same confidence level are shorter than those without AFC. Lastly, AFC improves the coverage for $\theta$ and $\sigma$.

\begin{center}
	\begin{table}[h]
		\centering
		\caption{ Inference Performance with AFC for Laplace Example.}
		\label{tab:Laplace2}%
		\begin{tabular*}{500pt}{@{\extracolsep\fill}lcccc@{\extracolsep\fill}}
			\toprule
			\textbf{Truth} & \textbf{Coverage} & \textbf{Expected CI Length} & \textbf{Expected Mean} & \textbf{Expected Median} \\
			\midrule
			$\theta=0$ & 0.870 & 0.4244& 0.0033 & 0.0032 \\ 
			$\sigma=1$ & 0.940 & 0.3466 & 0.9872 & 0.9796 \\
			\bottomrule
		\end{tabular*}
	\end{table}
\end{center}

\subsection{Nonlinear Data generating algorithm}\label{sec:nonlinear}
Consider the following non-linear model defined by the data generating algorithm
\[ \xb = {\mu} \times \bm{1} + \mu ^{\frac{q}{2}} \times \zb, \] 
where $\xb \in {{\mathbb R}}^m (m=3)$, $\mu \in {\mathbb R}$, and $\zb=(z_1,\ldots,z_m)$, where $z_i$ are realizations of independent standard normal random variables. Here, $\mu$ is the parameter of interest and $q/2$ is known. When $q=0$, this becomes a standard location parameter problem: $\bX =  {\mu} \times \bm{1} + \bZ $, $q=2$ is a scale parameter problem $\bX = \mu\times( \bm{1} + \bZ)$. and $q=4$ leads to a one-parameter exponential family. In this numerical experiment we choose $q=3$ in order to have a true non-linear model.

To train the FAE, we simulate 100,000 $\mu_k$ from the $Uniform(0,6)$ distribution and randomly split them to 80,000 training samples and 20,000 validation samples. We use an 11 layer fully-connected neural network as the encoder with ReLU as the activation function  \citep{nair2010rectified}  and with Adam as the optimizer \citep{kingma2014adam}. The FAE converges after 10 epochs. All FAEs in this paper are implemented using on Keras in Python \citep{chollet2015keras}.
%

\begin{figure}[t]
	\centering
	\includegraphics[width=0.75\linewidth]{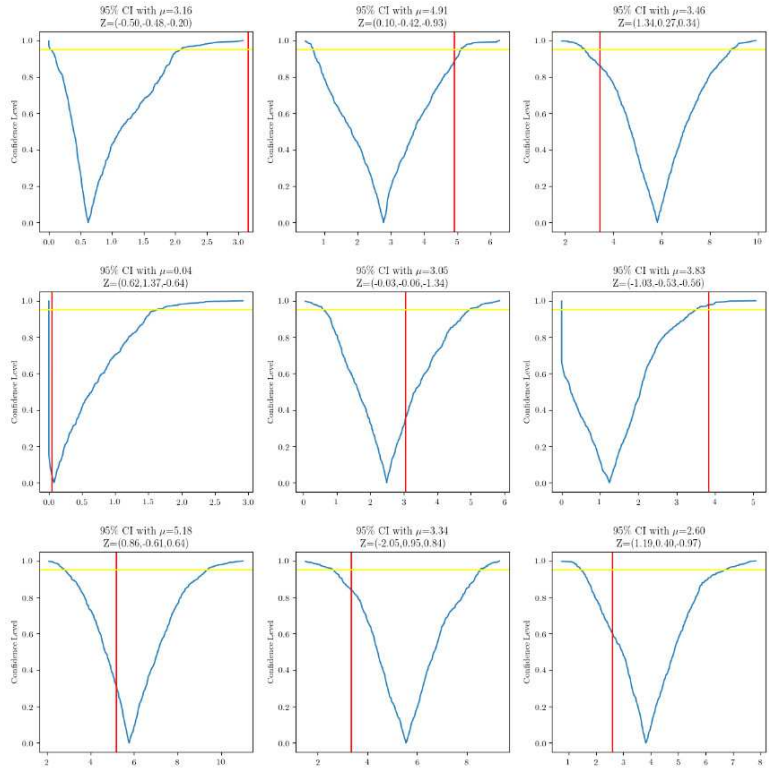}
	\caption{\label{EG1-9CC} Nine confidence curves for nonlinear data generating algorithm without AFC. The red vertical lines correspond to the true parameters. The intersection of the confidence level (yellow horizontal line) and the blue confidence curve shows the 95$\%$ confidence interval.}
\end{figure}

After the FAE has converged we will use the fitted encoder for inference. In particular we  first randomly sample $\bZ_j^\star, j=1,\ldots,n_{fid}$ ($n_{fid}=1000$) and use them together with the observed data $\xb$ in the encoder to obtain $n_{fid}$ samples of $\mu_j^\star$. Notice that the same observed value of $\xb$ is used for all  $\mu_j^\star$. 
Using the  $\mu_j^\star$ we can estimate an empirical distribution function of the GFD for $\mu$ and draw a corresponding confidence curve. Figure~\ref{EG1-9CC} shows confidence curves for nine different realizations of $\xb$. Most of the confidence intervals are wide and only 7 of 9 true $\mu$ are covered by the corresponding 95$\%$ level confidence interval. 

Table~\ref{tab:IP1} presents the results of a simulation study using FAE without AFC. In particular, we fixed four true $\mu \in {1,2,3,4}$, and observed 200 data sets  $\{ \xb_k,  k = 1, \ldots,200 \}  $ for each of the four values of $\mu$. For each $\xb_k$, we use 1,000 random $\bZ_j^\star$ to obtain $\mu_j^\star$ and use them to obtain fiducial mean, median and 90$\%$ confidence set. Table~\ref{tab:IP1} reports  {four key statistics averaged over the 200 data sets, namely, coverage, the expected length of the confidence intervals, the expected value of the fiducial mean and median }. We can see that the confidence sets are very wide and the expected fiducial mean and the expected fiducial median are biased. 
\begin{center}
	\begin{table}[h]
		\centering
		\caption{ Inference Performance without AFC for Nonlinear Data generating algorithm.}
		\label{tab:IP1}%
		\begin{tabular*}{500pt}{@{\extracolsep\fill}lcccc@{\extracolsep\fill}}
			\toprule
			\textbf{True $\mu$} & \textbf{Coverage} & \textbf{Expected CI Length} & \textbf{Expected Mean} & \textbf{Expected Median} \\
			\midrule
		1 & 0.985 & 2.03 & 1.07 & 0.90 \\ 
		2 & 0.905 & 3.50 & 2.64 & 2.49 \\
		3 & 0.865 & 4.25 & 3.85 & 3.81 \\
		4 & 0.870 & 4.28 & 4.48 & 4.45 \\
			\bottomrule
		\end{tabular*}
	\end{table}
\end{center}

\begin{figure}[t]
	\centering
	\includegraphics[width=0.7\linewidth]{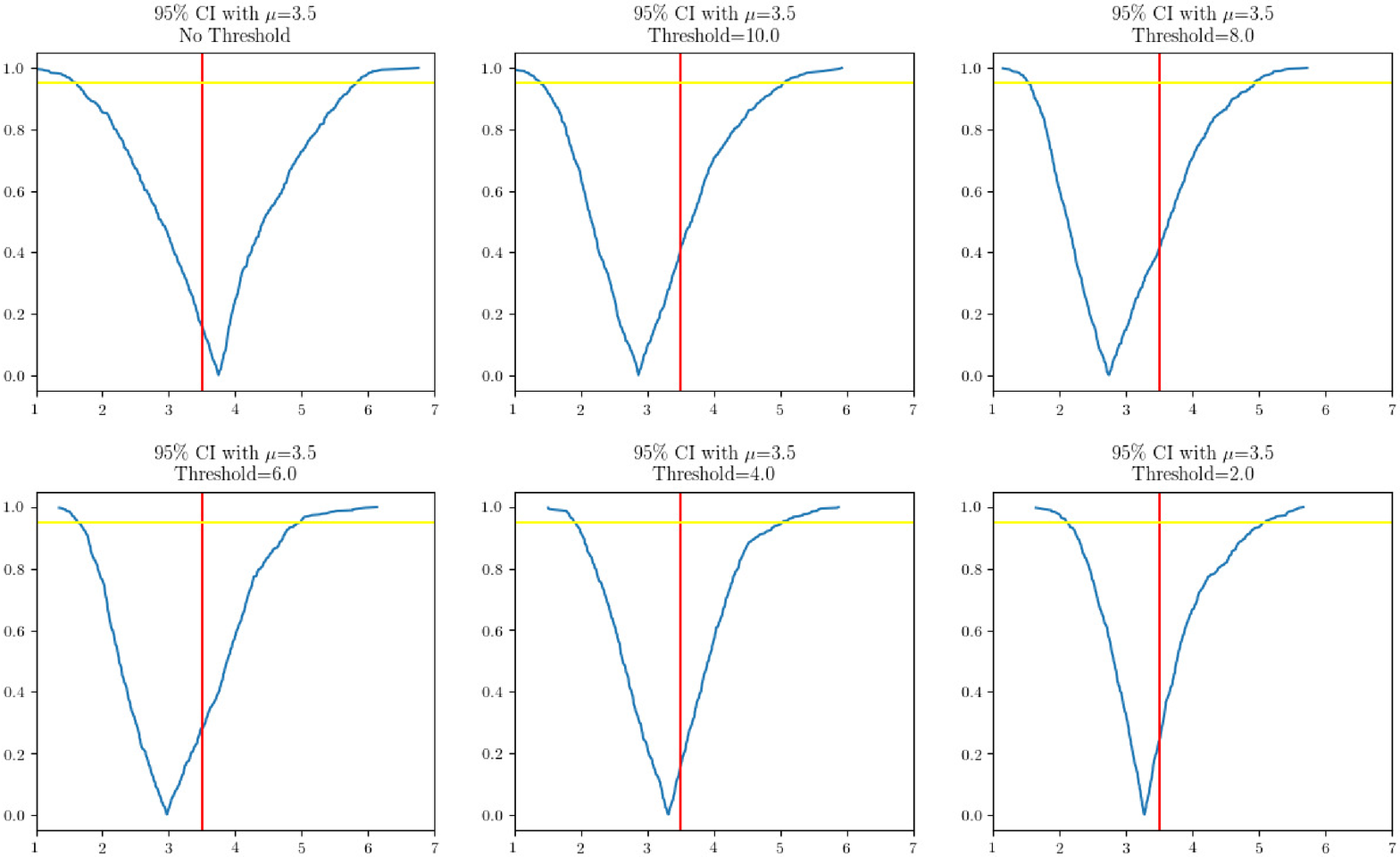}
	\caption{\label{EG1-Threshold} Confidence curves with different thresholds. The red vertical lines correspond to the true parameter ($\mu = 3.5$). }
\end{figure}

Next, we investigate the effect of AFC {for a fixed observation}. Figure~\ref{EG1-Threshold} shows confidence curves for the same data $\xb$ with different thresholds;  the true $\mu=3.5$.  As the threshold $\epsilon$ decreases, the marginal fiducial median is getting closer to the true $\mu$. Additionally, the fiducial distribution is becoming more concentrated. 

Finally, we repeated the simulation study for FAE with AFC to generate 1,000 threshold-admissible samples to form inference estimates. As shown in Table~\ref{tab:IP2}, the fiducial mean and median with AFC is less biased compared to the inference performance without AFC shown in Table~\ref{tab:IP1}. The empirical coverage is greater than the true 90$\%$ confidence level for all 4 settings. 
\begin{center}
	\begin{table}[h]
		\centering
		\caption{ Inference Performance with AFC for Nonlinear Data generating algorithm.}
		\label{tab:IP2}%
		\begin{tabular*}{500pt}{@{\extracolsep\fill}lcccc@{\extracolsep\fill}}
			\toprule
			\textbf{True $\mu$} & \textbf{Coverage} & \textbf{Expected CI Length} & \textbf{Expected Mean} & \textbf{Expected Median} \\
			\midrule
			1 & 0.95 & 3.10 & 1.44 & 1.06 \\ 
			2 & 0.95 & 4.07 & 2.55 & 2.18 \\
			3 & 0.97 & 3.43 & 3.22 & 2.99 \\
			4 & 0.94 & 3.30 & 3.98 & 3.89 \\
			\bottomrule
		\end{tabular*}
	\end{table}
\end{center}

\subsection{BOD}\label{dfi:s3:BOD}
Here we further test our AFC-corrected FAE solution on an non-linear regression algebraic model for {Biological Oxygen Demand} (BOD) \citep{bardsley_solonen_haario_laine_2014}
. 
The BOD model is often used for simulating the saturation of the growth of the observed response y that corresponds to a given variable x. The corresponding data generating algorithm is
\[\yb=f(\mu,\xb) = t_1(1 - e^{-t_2 \xb} ) + \zb\]
where the parameters $\mu=(t_1,t_2)$ are two scalars to estimate; the observed synthetic data have five design points $\xb=(2,4,6,8,10)$ and dependent observations $\yb=(0.152, 0.296, 0.413, 0.482, 0.567)$.
The observation errors $\zb=(z_1,\ldots,z_5)$ are assumed to be independent
and identically distributed mean zero Gaussian errors with the standard deviation $\sigma>0$. Following \cite{bardsley_solonen_haario_laine_2014} we fixed $\sigma = 0.015$.

\begin{figure}[t]
	\centering
	\includegraphics[width=0.8\linewidth]{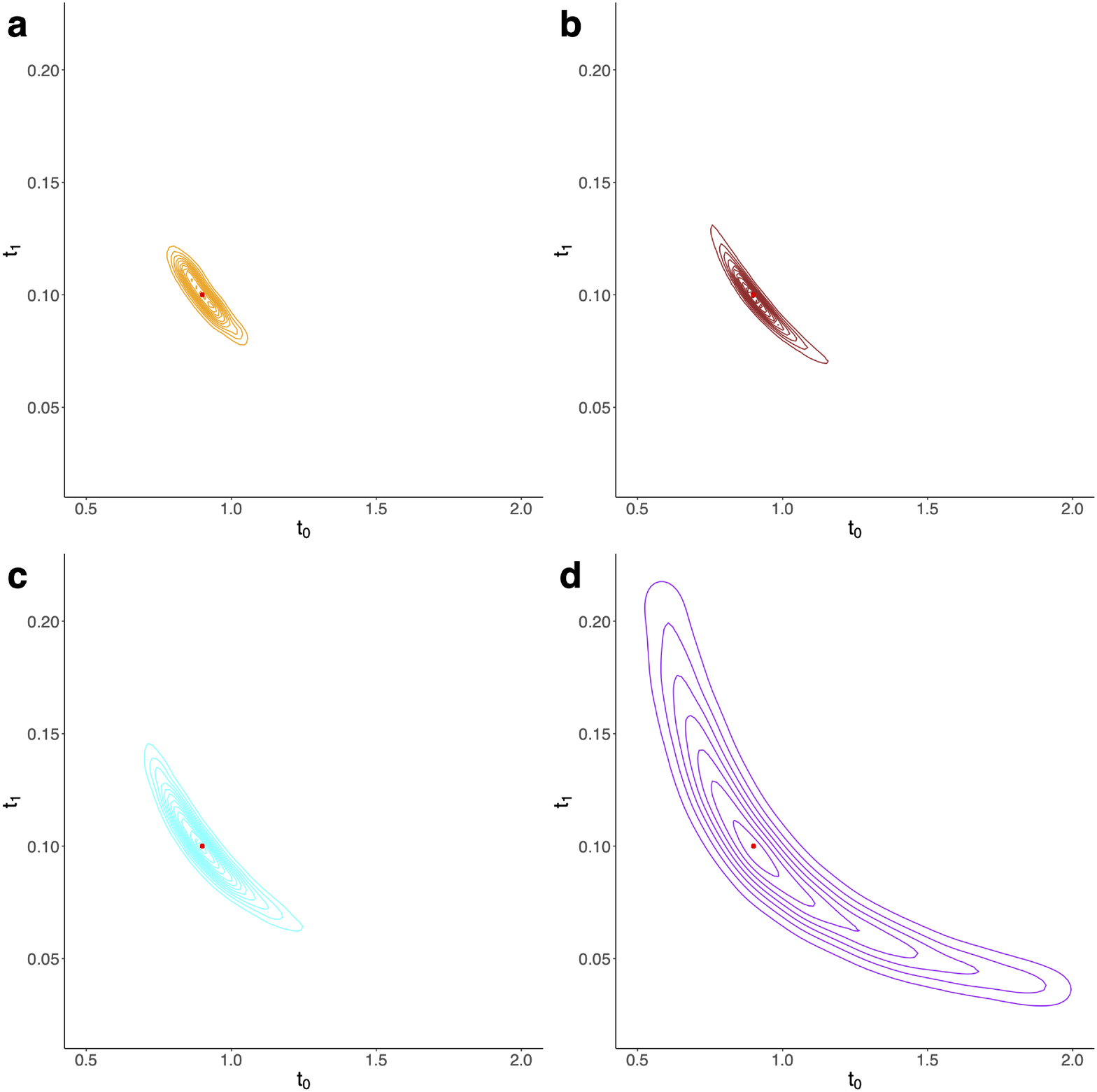}
	\caption{\label{EG3-2D}  Contour plots of the joint posteriors for the BOD example. (a) FAE estimation with AFC algorithm (orange); (b) fiducial estimation implemented  via HMC (brown); (c) parametric bootstrap (cyan); (d) {RTO} (Bayesian) solution (magenta). The red dots near the center of the contours represent the true parameters ($t_0=0.9, t_1=0.1$) that generated the observed data.}\end{figure}

\begin{figure}[t]
 	\centering
 	\includegraphics[width=\linewidth]{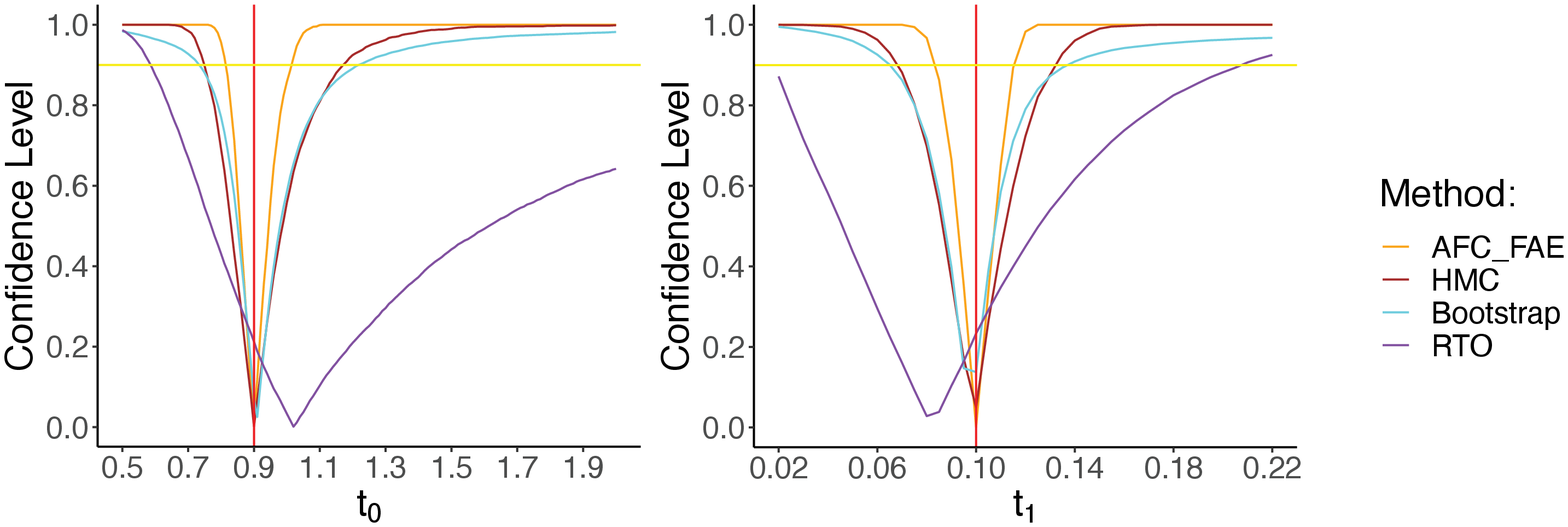}
 	\caption{\label{EG3-CC} Confidence curves of marginal distributions for the BOD example.  The  colors associated with the methods are the same as in Figure~\ref{EG3-2D}. The red vertical lines are the true parameters ($t_0=0.9, t_1=0.1$) generated the observed data. The yellow horizontal line indicates the 90\% confidence intervals.}
 \end{figure}

We used four different methods to analyze the BOD data with the resulting distributions reported in Figure~\ref{EG3-2D}. The FAE with AFC is in panel a); the GFD using Theorem~\ref{Jacobian} implemented using Hamiltonian Monte Carlo in STAN \citep{RStan} is in panel b); the parametric bootstrap  \citep{tibshirani1993introduction} is in panel c); and the Bayesian solution of \cite{bardsley_solonen_haario_laine_2014} using Metropolis-Hastings algorithm with a simplified fiducial-like proposal, {called randomize-then-optimize (RTO), }
is in panel d).
Notice that all methods yield a "banana-shaped" distribution centered around the true parameters ($t_1=0.9, t_2=0.1$), indicated by the red dots in Figure \ref{EG3-2D}. The AFC-corrected FAE solution (orange contour) achieve the thinnest banana-shape among all, while the RTO solution is the widest. Furthermore, we report confidence curves based on the marginal distributions of the two parameters separately in Figure~\ref{EG3-CC}. Again our AFC-corrected FAE solution is the thinnest among the four methods. Both the thinnest banana shape and confidence curve indicate that our AFC-corrected FAE solution is the most concentrated near the truth. 

In summary, we have shown how to use AFC with and without the analytical form of the inverse function, and by implementing AFC we improved the inference performances of GFI. Our FAE provides an accurate approximation of the inverse function. 

\section{Discussion}\label{sec5}
In this paper, we first proposed a fiducial autoencoder for the circumstance in which the analytical form of the inverse function is not available or the marginal fiducial density is intractable. The universal approximation theorem provides theoretical guarantees for the approximation performance of our FAE, and our simulations further validate our approach. The proposed FAE can accurately approximate the inverse function, and it can be efficiently combined with the AFC algorithm to provide valid and accurate inferences of the true parameters. 
The AFC algorithm is similar to ABC and provides an insight into the relationship between Bayesian and fiducial distribution; the use of a prior versus solving an optimization problem when proposing $\mu$. 

For modern machine learning and deep learning communities, data are usually implicitly modeled, while for fiducial inference, we explicitly model the modeling mechanism behind the data through our data generating algorithm. Under the FAE framework, we are combining those two approaches: we incorporate the data generating algorithm as a decoder (explicitly model the forward process); we keep deep neural network (implicit model) as an encoder to computationally solve the inverse problem. Furthermore, FAE might help the deep learning community to understand the neural work through our fiducial autoencoder. For example, a neural network is often regarded as a non-linear regression tool, but one main difficulty for DL is that the true function is unknown. When the inverse function has an analytical solution, we can quantify the biases in a certain sense. Thus studying the theoretical properties of FAE could potentially be helpful to understand the neural network.


One limitation of our AFC-corrected FAE solution is that the inference results are sensitive to the training data FAE learned from. Neural networks are known for their ability to perform well on the similar data it has seen before. If most of the parameters of our simulated training data are far from the true parameter of the observed data, or FAE does not see enough similar training data as the true parameters of the observed data, FAE, even with AFC, might provide a biased estimation. A simple remedy for that is to generate the training data around the least square estimation of the parameter. In other words, we only require the FAE to learn the inverse function around the true parameter, instead of the whole parameter space, which can be much harder. 

We have focused on cases with completely known data-generating algorithm and our pre-defined neural network architecture (fully connected neural networks) in the numerical experiments. The main purpose of these studies is to demonstrate the approximation performance of the FAE and the validity of the AFC algorithm. Thus we did not tune the best neural network architecture for the encoder part of the FAE. Note we do not have a specific
requirement for the network structure of encoder. Any standard deep neural networks can be used to construct an encoder. However, to learn a complex inverse function inevitably requires a more sophisticated construction of encoder. This can be a potential issue, but with the rapid development of the deep learning communities, we expect our FAE would be of great use with elegant construction. The competitive performance of AFC corrected FAE solution both in terms of efficiency and accuracy suggests that this is a promising area for future research.

\section*{DATA AVAILABILITY STATEMENT}\label{sec6}

Python code for the simulation results is available from the authors upon request.


%
%
\bibliographystyle{wb_stat}
\bibliography{yourbibfile}%

\end{document}